\documentclass[doublecol]{epl2}
\usepackage{amsmath}
\usepackage{amssymb}

\title{Brillouin-Wigner perturbation theory in open electromagnetic
\mbox{systems}}

\shorttitle{ Brillouin-Wigner perturbation theory in open
electromagnetic systems}

\author{E.\,A. Muljarov\inst{1} \and W. Langbein\inst{1} \and R. Zimmermann\inst{2}}
\shortauthor{E.\,A. Muljarov \etal}

\institute{
  \inst{1} School of Physics and Astronomy,
Cardiff University, The Parade, CF24\,3AA, Cardiff, United Kingdom\\
  \inst{2} Institut f\"ur Physik der Humboldt-Universit\"at zu Berlin,
Newtonstrasse 15, D-12489 Berlin, Germany }

\pacs{03.50.De}{Classical electromagnetism, Maxwell equations}

\pacs{42.25.-p}{Wave optics}

\pacs{{03.65.Nk}}{{Scattering theory}}

\pacs{41.20.-q}{Applied classical electromagnetism}

\abstract{ A Brillouin-Wigner perturbation theory is developed for
open electromagnetic systems which are characterised by discrete
resonant states with complex eigenenergies. Since these states are
exponentially growing at large distances, a modified normalisation
is introduced that allows a simple spectral representation of the
Green's function. The perturbed modes are found by solving a
linear eigenvalue problem in matrix form. The method is
illustrated on exactly solvable one- and three-dimensional
examples being, respectively, a dielectric slab and a
microsphere.}

\begin{document}

\maketitle

Eigenvalue problems are often not analytically solvable, and
approximate solution schemes are employed, which use a known
(ideally analytic)  solution of a similar problem, differing from
the problem of interest by the `perturbation'. The most well known
scheme is the Rayleigh-Schr\"odinger perturbation
theory~\cite{Shroedinger26}, which writes the solution as series
expansion in terms of the perturbation involving sums over all
eigenstates, and thus introduces the notion of perturbation
orders. However, the convergence properties of such series limit
the treatable perturbation strength, and the mathematical
complexity limits the expansion typically to a few orders. One way
to overcome this limitation is to truncate the infinite complete
set of unperturbed eigenstates to a finite number and then use
matrix diagonalisation algorithms to treat the perturbation within
this basis exactly, {\it i.e.} to all orders. In quantum
mechanics, this method is called Brillouin-Wigner (BW)
perturbation theory~\cite{Brillouin32}. If the unperturbed
spectrum contains bound states only (as for the harmonic
oscillator), it allows to treat the effect of perturbation to any
required accuracy by taking a sufficiently large number of states.
However, if a continuum of scattering states is important, the use
of the BW theory becomes problematic.  In electromagnetic systems,
such a continuum is often the dominating part of the spectrum and
discrete bound states (evanescent waves) are the exception.

Both Rayleigh-Schr\"odinger and BW perturbation theories are well
established when treating Hermitian problems which describe
conservative systems having real-valued eigenenergies and
orthogonal and normalised wave functions. In open systems instead,
an eigenstate acquires a certain linewidth which characterises its
finite lifetime, so that the eigenenergy becomes
complex-valued~\cite{Gamow28}. The corresponding eigenfunctions
are not orthogonal and grow exponentially in the outer space so
that they cannot be normalised by integrating their square
modulus~\cite{Zeldovich60,Zeldovich69,Weinstein69}. Both the
energies and the wave vectors of these states have negative
imaginary parts resulting in a wave function
$\propto\exp[-\Gamma(t-r/v)]$ ($\Gamma=- {\rm Im}\,\omega$, $v$ is
the phase velocity). The latter expression provides a physical
picture of the exponential growth as a wave front propagating away
from the system which has been excited at an earlier time. In this
way stationary bound states become
quasi-stationary~\cite{Zeldovich69} in open systems due to leakage
to the outside. In the literature these states are also known as
decaying states~\cite{More71}, resonant states (RS)~\cite{More73},
leaky modes~\cite{Olshansky79}, quasi-guided
modes~\cite{Tikhodeev02}, quasi-normal
modes~\cite{Leung94,Leung96}, etc. Moreover, going away from the
real-energy axis into the complex energy plane, a continuum of
stationary scattering states, which is problematic in perturbation
theory, can be effectively eliminated from the spectrum and
replaced by a countable number of RS~\cite{Weinstein69}. Formally,
this is achieved by imposing boundary conditions of no incoming
waves, {\it i.e.} no waves travelling towards the system.

The Rayleigh-Schr\"odinger perturbation theory of RS has been
developed for quantum-mechanical~\cite{More71} and electromagnetic
systems~\cite{Leung96}. For weak perturbations, an approximate BW
theory limited to degenerate modes only has been
reported~\cite{Lai90}. Also, an exact method of calculation of RS
perturbed by a point scatter has recently been
developed~\cite{Dettmann08}. However, sharp resonances in
three-dimensional structures cannot be calculated by known methods
in electrodynamics. Thus it is especially important to develop a
BW perturbation theory, which could treat such resonant states
under a perturbation of arbitrary strength and shape. Such a
theory was only partially formulated in electrodynamics more than
40 years ago\cite{Weinstein69} and later on proposed also in the
quantum-mechanical context~\cite{Bang78,Lind93}, but has not been
completed neither applied to numerically solve relevant problems.

In this Letter we present a BW perturbation theory of open
electromagnetic systems and validate it on exactly solvable
one-dimensional (1D) and three-dimensional (3D) examples, for both
weak and strong perturbations. Although the RS have complex
energies and wave functions exponentially increasing outside the
system~\cite{Zeldovich69}, orthogonality and normalisation can be
established. Then, the expansion of the perturbed system into the
RS of unperturbed system can be reduced to the standard BW form of
diagonalisation of the sum of a diagonal matrix of the unperturbed
system and a perturbation matrix.

To make the idea clear and the derivation transparent we present
our BW theory first on an effective 1D problem taking a
homogeneous dielectric slab as unperturbed system and then, after
bringing it to a general form, use it for a dielectric
microsphere.

We consider a dielectric slab in vacuum with thickness $2a$ which
is described by a real dielectric constant along the spatial
coordinate $z$ [see the inset in fig.\,1(a)]:
 \begin{equation}
\varepsilon(z)=\left\{
 \begin{array}{cc}
\varepsilon_s\,, &|z|<a\,,\\
1\,, & |z|\geqslant a\,.
 \end{array}
 \right.
 \end{equation}
The permeability is $\mu=1$ everywhere throughout this work. For
transversal eigenmodes ${\bf E}_n=\hat{{\bf x}} E_n/\sqrt{S}$ ($S$
is the normalisation area in the $xy$-plane) having zero in-plane
wave number, $k_x=k_y=0$, and time-dependent part given by
$\exp(-ic k_n t)$ with the complex frequency $c k_n$, the Maxwell
equation is reduced to
 \begin{equation}
\left[\frac{d^2}{dz^2} + \varepsilon(z) k_n^2\right]E_n(z)=0
 \label{ME0}
 \end{equation}
with the electric field $E_n$ changing smoothly {({\it i.e.}
having a continuous derivative)} across the boundaries at $z=\pm
a$. Assuming the boundary conditions (BC) of only outgoing waves
results in the eigenfunctions
 \begin{equation}
E_n(z)=\left\{
 \begin{array}{lc}
(-1)^n A_n e^{-ik_n z}\,, &z<-a\\
B_n\bigl[e^{i\sqrt{\varepsilon_s}k_n z}+(-1)^n e^{-i\sqrt{\varepsilon_s}k_n z}\bigr]\,, & |z|<a\\
A_n e^{ik_n z}\,, & z>a
 \end{array}
 \right.
 \label{modes}
 \end{equation}
and the wave number eigenvalues
 \begin{equation}
k_n=\frac{1}{2a\sqrt{\varepsilon_s}}\left(\pi n
-i\,\ln\alpha\right),\ \ \ n=0,\,\pm1\,,\pm2\,\dots,
 \label{1Dspectrum}
 \end{equation}
where  $\alpha=(\sqrt{\varepsilon_s}+1)/(\sqrt{\varepsilon_s}-1)$.
 The eigenfunctions $E_n$
are orthogonal and normalised in a well-known
way~\cite{Weinstein69}
 \begin{equation}
\int_{-a}^a\varepsilon_s E_n E_m dz -\frac{E_n(a)
E_m(a)+E_n(-a)E_m(-a)}{i(k_n+k_m)}=\delta_{nm}
 \label{norm1D}
 \end{equation}
which is determining the amplitudes to
 \begin{equation}
A_n=\frac{e^{-ik_n a}}{\sqrt{a(\varepsilon_s-1)}}\,,\ \  \ \ \
B_n=\frac{(-i)^n}{2\sqrt{a\varepsilon_s}}\,.
 \end{equation}
Let us also construct the unperturbed Green's function (GF)
$G(z,z';k)$ which satisfies the equation
 \begin{equation}
\left[\frac{d^2}{dz^2} + \varepsilon(z)
k^2\right]G(z,z';k)=\delta(z-z')\,,
 \label{GFequ}
 \end{equation}
and obeys the same BC of no incoming waves. {As in the case of a
square-barrier potential in quantum mechanics~\cite{Aguiar93},
this GF can be calculated analytically.} Inside the slab
($-a<z,z'<a$) {it} has the following explicit form:
 \begin{eqnarray}
G(z,z';k)&=&\frac{1}{2iq}\,\frac{e^{iq(a-z_>)}+\alpha
e^{-iq(a-z_>)}}{e^{iqa}+\alpha e^{-iqa}}\nonumber\\
&&\times \frac{e^{iq(a+z_<)}+\alpha
e^{-iq(a+z_<)}}{-e^{iqa}+\alpha e^{-iqa}}
 \end{eqnarray}
where $z_>=\max(z,z')$, $z_<=\min(z,z')$, and
$q=k\sqrt{\varepsilon_s}$. The analytic continuation of the GF
into the complex $k$-plane has simple poles at $k=k_n$ and $k=0$.
Then, {similar to the resolvent expansion of a Hermitian operator
in the energy plane~\cite{Sukumar90}}, it can be written, using
the Mittag-Leffler theorem, as a sum of partial
fractions~\cite{Newton60,More71}:
\begin{equation}
G(z,z';k)=\sum_n\frac{E_n(z)E_n(z')}{2k_n(k-k_n)}+\frac{1}{2ik}
=\sum_n\frac{E_n(z)E_n(z')}{2k(k-k_n)}
 \label{GFseries}
 \end{equation}
where the residua were expressed in terms of the normalised
eigenmodes eq.(\ref{modes}) and a sum rule $\sum_n
E_n(z)E_n(z')/k_n=i$ following from the high-frequency asymptotics
of the GF~\cite{Bang81} was used.

Let us now consider an arbitrary real perturbation
$\Delta\varepsilon(z)$ of the dielectric constant {\it inside} the
slab. The wave equation for the perturbed resonant states ${\cal
E}_\nu$ and wave numbers $\varkappa_\nu$ is given as in
eq.\,(\ref{ME0}) by
 \begin{equation}
\left\{\frac{d^2}{dz^2} +
\bigl[\varepsilon(z)+\Delta\varepsilon(z)\bigr]
\varkappa_\nu^2\right\} {\cal E}_\nu(z)=0
 \label{ME}
 \end{equation}
with BC ${\cal E}_\nu(z) \propto \exp(i \varkappa_\nu |z|)$ at
$|z|>a$. The new GF ${\cal G}(z,z';k)$ is related to the
unperturbed one via the Dyson equation
 \begin{eqnarray}
 &&\hskip-10mm {\cal G}(z,z';k)=G(z,z';k)\nonumber\\
 &&-k^2\int_{-a}^a G(z,z'';k)\Delta\varepsilon(z'')
 {\cal G}(z'',z';k)dz''.
 \label{Dyson}
 \end{eqnarray}
On the other hand, a spectral representation of the perturbed GF,
similar to eq.\,(\ref{GFseries}), can be used. It is derived in
the Appendix for an arbitrary profile of the dielectric function
inside the slab. Then, following ref.\,\cite{More71} we equate
residua at the poles $k=\varkappa_\nu$ in eq.\,(\ref{Dyson}). This
results in the following relationship between the unperturbed and
perturbed modes:
 \begin{equation}
{\cal E}_\nu(z) = \sum_n\frac{E_n(z)\int_{-a}^a
E_n(z')\Delta\varepsilon(z'){\cal
E}_\nu(z')dz'}{2(k_n/\varkappa_\nu-1)}\,.
 \label{Enewold}
 \end{equation}
In the interior region $|z|<a$ which contains the perturbation,
the perturbed RS can be expanded into the unperturbed ones,
exploiting the completeness of the latter,
 \begin{equation}
{\cal E}_\nu(z) = \sum_n b_{n\nu} E_n(z)\,.
 \label{E-expansion}
 \end{equation}
Substituting this expansion into eq.\,(\ref{Enewold}) and equating
coefficients at the same basis functions $E_n(z)$ results in the
following matrix equation:
 \begin{equation}
2b_{n\nu}=\frac{1}{k_n/\varkappa_\nu-1}\sum_m V_{nm} b_{m\nu}\,.
 \end{equation}
Introducing new coefficients $c_{n\nu}=b_{n\nu}\sqrt{k_n}$ to
symmetrise the matrix problem, we arrive at
 \begin{equation}
\sum_m\left(\frac{\delta_{nm}}{k_n}+\frac{V_{nm}}{2\sqrt{k_n
k_m}}\right) c_{m\nu}=\frac{1}{\varkappa_\nu} c_{n\nu}\,.
 \label{BWtheory}
 \end{equation}
Equation~(\ref{BWtheory}) is the central result in our BW
perturbation theory of resonant states. Similar to the BW theory
of bound states in quantum mechanics, eq.\,(\ref{BWtheory})
represents a {\em linear eigenvalue problem} which requires
diagonalisation of a matrix depending only on the unperturbed
spectrum and the perturbation~\cite{footnote-QM}. However, instead
of real eigenenergies and a Hermitian perturbation matrix
characteristic for conservative systems, we are dealing here with
complex-valued bare and perturbed eigenvalues, $1/k_n$ and $
1/\varkappa_\nu$, respectively, and a complex symmetric
perturbation matrix $V_{nm}/(2\sqrt{k_n k_m})$, where
 \begin{equation} V_{nm}=\int
\Delta\varepsilon({\bf r}){\bf E}_n({\bf r})\cdot{\bf E}_m({\bf
r})\,d {\bf r}\,.
 \label{pert}
 \end{equation}
{ While from the formulation of the method both the unperturbed
system and the perturbation $\Delta\varepsilon({\bf r})$ can be
located inside any finite area, we found that good convergency is
achieved for a perturbation within the area of the unperturbed
system. }

To illustrate the accuracy of the method, we use as unperturbed
system the slab with $\sqrt{\varepsilon_s}=1.5$. The unperturbed
discrete spectrum is given by eq.\,(\ref{1Dspectrum}) and shown in
fig.\,1\,({\it a,b}) by circles with a dot. We introduce a strong
non-symmetric perturbation $\Delta\varepsilon=10$ and 100 in the
layer $(a/2<z<a)$ [see the inset in fig.\,1(a)], thus increasing
in the latter case the refractive index by an order of magnitude.
The exact eigenvalues $\varkappa^{\rm exact}_\nu$ are given in
the Appendix by eq.\,(A.9). %\ref{EVapp}).
They are compared in
fig.\,1\,({\it a,b}) (squares) with our numerical result (crosses)
calculated according to eq.\,(\ref{BWtheory}) where the sums have
been truncated at $-N \le n \le N$, yielding a matrix dimension of
$(2N+1)$. The relative error $|\varkappa_\nu/\varkappa^{\rm
exact}_\nu-1|$ is quantified in fig.\,1\,({\it c,d}) for different
$N$. States close to the truncation border ({\it i.e.} with large
values of $|\nu|$) are always poorly calculated. However, as the
matrix size increases, a state with a given $\nu$ is moving closer
to the middle of the matrix, so that its relative error decreases
as $ N^{-3}$. The exact eigenfunctions (the electric field ${\cal
E}_\nu$) are reproduced to a similar accuracy.

\begin{figure}[t]
\onefigure[width=8.5cm]{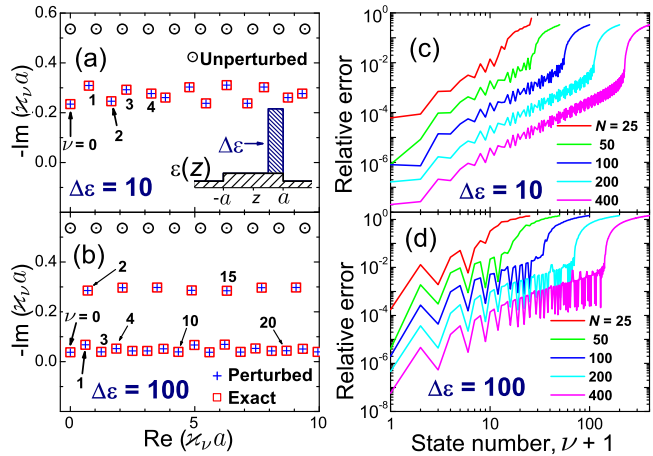}
 \caption{ (colour online)
Resonant states of a dielectric slab. Exact and calculated via
eq.\,(\ref{BWtheory}) wave numbers $\varkappa_\nu$ ({\it a,b})
along with the relative error of the calculation ({\it c,d}), for
different truncation $N$ of the matrix problem. Unperturbed: RS of
a dielectric slab in vacuum (black circles with a dot). The slab
with $\varepsilon_s=2.25$ occupies the area $-a<z<a$. Perturbed:
Exact (red rectangles) and calculated  (blue crosses) RS wave
numbers $\varkappa_\nu$ of the same slab but with a higher
refractive index in the layer $(a/2<z<a)$, the change there being
$\Delta\varepsilon=10$ ({\it a,c}) or $\Delta\varepsilon=100$
({\it b,d}). Inset: profiles of the dielectric constant of the
unperturbed  and  perturbed systems.
 }
\end{figure}

Writing the perturbation in the form of eq.\,(\ref{pert}), the
bare eigenmodes ${\bf E}_n({\bf r})$ are supposed to be normalised
as given by eq.\,(\ref{norm1D}) in the 1D case. In
quantum-mechanical systems, normalisation of RS was discussed {\it
e.g.} in refs.~\cite{More71,More73,Watson86}. A similar approach
was used to normalise eigenmodes in spherically symmetric open
electromagnetic systems~\cite{Weinstein69,Lai90,Leung96}. Some
important details of such normalisation are given in the Appendix
along with the spectral representation of the Green's function,
see eqs.\,(A.23) and (A.24).
%eqs.\,(\ref{wn}) and (\ref{GFapp}).
Below we show, however, how this normalisation can be done in the
general 3D case.

The electric field modes ${\bf E}_n$ obey Maxwell's equation
 \begin{equation}
 \nabla\times\nabla\times{\bf E}_n({\bf r})=k^2_n\varepsilon({\bf r}){\bf
E}_n({\bf r})\,,
 \label{me3D}
\end{equation}
 where the dielectric constant $\varepsilon({\bf r})$ is assumed
to deviate from unity only inside a sphere of radius
$a$. %~\cite{note1}.
The perturbation $\Delta\varepsilon({\bf r})$
is located within the same sphere. Multiplying eq.\,(\ref{me3D})
by ${\bf E}_m({\bf r})$ and integrating the dot product over the
sphere of radius $R\geqslant a$ by parts, we obtain the {\em
orthogonality} condition
 \begin{eqnarray}
0&=&(k_n^2-k_m^2)\int _R d{\bf r}\varepsilon({\bf r}){\bf
E}_n({\bf r})\cdot{\bf E}_m({\bf r})\nonumber\\
&&-\int _R dS \left({\bf E}_n\cdot\frac{\partial{\bf
E}_m}{\partial r}-{\bf E}_m\cdot\frac{\partial{\bf E}_n}{\partial
r}\right),
 \label{orthog}
 \end{eqnarray}
where the last integral is taken over the surface of the
$R$-sphere. Both integrals in eq.\,(\ref{orthog}) are
$R$-dependent and divergent at $R\to\infty$. There is, however, an
exact cancellation of these $R$-dependencies thus removing the
divergencies~\cite{note2}. In a degenerate case $k_n=k_m$ ($n\neq
m$), both the volume and the surface integrals vanish by symmetry
showing the orthogonality.

\begin{figure}[t]
\onefigure[width=8.5cm]{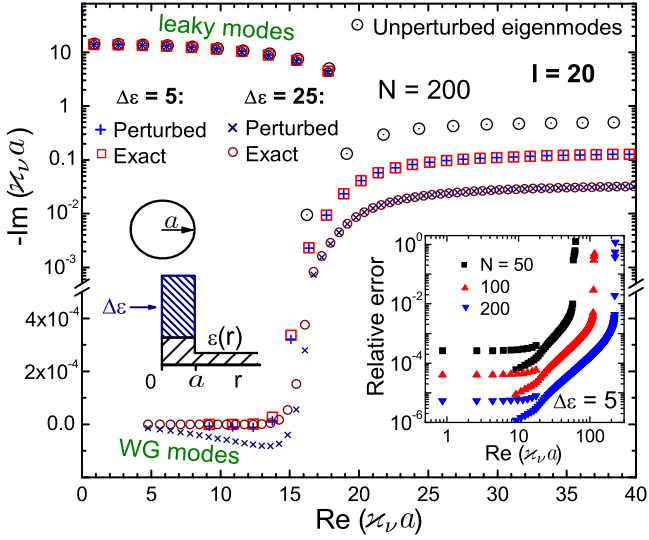}
 \caption{ (colour online)
Resonant states of a sphere, with the angular momentum $l=20$ (TE
modes). Unperturbed: Exact RS wave numbers $k_n$ of a homogeneous
dielectric microsphere in vacuum, with radius $a$ and refractive
index 1.5 (black circles with a dot). Perturbed: Wave numbers
$\varkappa_\nu$ of a microsphere with the same radius $a$ but
increased dielectric constant ($\Delta\varepsilon=5$: red
rectangles and blue crosses +; $\Delta\varepsilon=25$: red circles
and blue crosses $\times$). Results of eq.\,(\ref{BWtheory}) are
compared with exact values. Left inset: profiles of the dielectric
constant of the unperturbed and perturbed systems. Right inset:
relative errors of calculation with $\Delta\varepsilon=5$ for
different truncation $N$ of the matrix problem. }
\end{figure}

Equation~(\ref{orthog}) is now used as a starting point for the
eigenstate normalisation. In the outside range  $r>a$, where
$\varepsilon({\bf r})=1$, we take advantage of the functional
behaviour ${\bf E}_n({\bf r})={\bf F}_n(k_n{\bf r})$ with an
auxiliary function ${\bf F}_n$ being a superposition of vector
spherical harmonics~\cite{Stratton}. The analytic continuation
${\bf E}(k,{\bf r})$ of the eigenfunction ${\bf E}_n({\bf r})$
around the point $k_n$ in the complex $k$-plane is introduced in
such a way that ${\bf E}(k,{\bf r})={\bf F}_n(k{\bf r})$ and
satisfies eq.\,(\ref{me3D}) (with $k_n$ being replaced by the
$k$). The {\em normalisation} of ${\bf E}_n({\bf r})$ is then
defined as follows:
 \begin{eqnarray}
1&=&\int _R d{\bf r}\varepsilon({\bf r}){\bf E}_n^2({\bf r})
\nonumber\\
 && -\lim_{k\to k_n}\frac{\displaystyle \int _R dS
\left({\bf E}_n\cdot\frac{\partial{\bf E}}{\partial r}-{\bf
E}\cdot\frac{\partial{\bf E}_n}{\partial r}\right)}{k_n^2-k^2}\,.
 \label{normaliz}
 \end{eqnarray}
It is again $R$-independent, so that eq.\,(\ref{normaliz}) can be
evaluated for $R=a$. Furthermore, due to the above mentioned
functional behaviour, the limit $k\to k_n$ can be taken explicitly
reducing the last term in eq.~(\ref{normaliz}) to
 \begin{equation}
+\frac{1}{2k_n^2}\int_a dS \left[{\bf E}_n\cdot\frac{\partial {\bf
E}_n}{\partial r}+r{\bf E}_n\cdot\frac{\partial^2 {\bf
E}_n}{\partial r^2}-r\left(\frac{\partial {\bf E}_n}{\partial
r}\right)^2\right].
 \end{equation}

To illustrate the 3D case, we take as unperturbed system a
dielectric microsphere of radius $a$ and refractive index equal to
1.5. We consider a perturbation (sketched in the left inset of
fig.\,2) which uniformly increases the refractive index inside the
sphere ($\Delta\varepsilon=5$ and 25) leaving its radius
unchanged. Since both the unperturbed and perturbed systems have
spherical symmetry, the angular momentum $l$ is a good quantum
number and can be used to classify the RS. Additionally,
eigenmodes can be classified as transverse electric (TE) and
transverse magnetic (TM). These TE and TM modes are not mixed by
our perturbation. The dielectric microsphere has a rich spectrum
containing both leaky and high-Q whispering gallery (WG) modes,
the latter having a small imaginary part of $\varkappa_\nu$. The
difference between these two types of modes becomes more prominent
at larger values of $l$, see the spectrum for $l=20$ in fig.\,2.
For TE modes, the exact solutions (squares and circles) are well
reproduced by our method (crosses). To calculate the eigenstates
shown in fig.\,2, the matrix problem was truncated to $N =200$.
The accuracy quickly improves as the matrix dimension (2$N$)
increases, again with a relative error $\propto N^{-3}$, as is
seen in the right inset of fig.\,2.

In conclusion, a Brillouin-Wigner perturbation theory of resonant
states in open electromagnetic systems has been developed and
validated on exactly solvable 1D and 3D examples. Although the
wave functions are exponentially growing at large distances and
thus cannot be normalised in the usual way, the eigenvalue problem
is reduced to the standard linear form, typical for conservative
systems. Practically, it only requires diagonalisation of a
complex symmetric matrix which consists of the bare spectrum and
the perturbation. The simple spectral representation of the
Green's function in terms of the calculated perturbed resonant
states allows an easy access to measurable quantities such as
emission, scattering, and transmission.

\acknowledgments

The authors thank {\scshape P. Borri}, {\scshape A.\,L. Ivanov}
and {\scshape L. Chantada} for valuable discussions. E.A.M.
acknowledges financial support of WIMCS. W.L. acknowledges support
from BBSRC, grant BB/E005624.
\bigskip

{\scshape Appendix}

\section{Spectral representation of the Green's function in 1D}
The central quantity for calculating {\it e.g.} transmission and
emission properties of a dielectric system is the corresponding GF
which in 1D satisfies eq.\,(\ref{GFequ}) with an arbitrary profile
$\varepsilon(z)$ of a dielectric slab embedded in vacuum [{\it
i.e.} $\varepsilon(z) = 1$ for $|z|>a$].
It can be solved in the form
%
%\begin{equation} \label{GFWron}
$$
G(z,z';k) = \frac{E_L(z_<,k)\, E_R(z_>,k)}{W_{L,R}(k)}
 \eqno{\rm (A.1)}
$$
%\end{equation}
%
with $E_L(z,k)$ and $E_R(z,k)$ being the left- and
right-exponential homogeneous solutions,
%
%\begin{eqnarray}
$$
E_L(z,k)  = e^{-i k z} \;\;\mbox{for}\;\; z<-a \, ,\nonumber \\
$$
$$
E_R(z,k)  =  e^{+i k z} \;\;\mbox{for} \;\;z>+a \, ,
 \eqno{\rm (A.2)}
$$
% \label{ELR}
%\end{eqnarray}
so that the GF obeys the BC of outgoing waves at either side, for
real positive wave number $k$. The Wronskian $W_{L,R}(k) \equiv
E_L(z,k) E_R'(z,k) - E_L'(z,k) E_R(z, k)$ in the denominator takes
care of the delta inhomogeneity in eq.\,(\ref{GFequ}).

The analytic continuation of the GF into the complex $k$-plane has
simple poles at $k_n$  where the Wronskian vanishes, {\it i.\,e.}
where left- and right-exponential solutions agree (up to a
factor). This defines the eigenmodes $ E_n(z) \propto E_L(z, k_n)
\propto E_R(z, k_n)$. They have an exponential behavior on both
sides of the slab,
%
%\begin{equation}
$$
E_n(z) \propto \exp(i k_n |z|) \;\; \mbox{for} \; |z|>a \, .
 \eqno{\rm (A.3)}
$$
%\end{equation}
%
For any {\it real} dielectric function, the eigenstates come in
pairs [$E_n(z)$ and $E^*_n(z)$], giving rise to wave numbers which
are positioned symmetrically with respect to the imaginary $k$
axis.
%\revision{X}
%If $|{\rm Im}\,k_n|$ is small, the GF and other quantities
%like transmission exhibit a sharp structure around $k \approx {\rm
%Re}\,k_n$, with a width being related to $|{\rm Im}\,k_n|$. The
%case Im$\,k_n >0$ corresponds to bound states which are
%exponentially decaying (evanescent) outside the slab and have
%Re$\,k_n=0$. These states are orthogonal and normalised in the
%usual way.  In 1D problems such states can appear either due to
%regions with negative dielectric constant, or for non-normal
%incidence on a slab. They are quite analogous to quantum well
%localized states in quantum-mechanical problems.
{The states} with Im$\,k_n <0$ which we call RS increase
exponentially outside the slab and therefore the orthogonality
relation and normalisation have to be modified.

The orthogonality is established by integrating Maxwell's equation
(\ref{ME0}) for $E_n(z)$, multiplied with $E_m(z)$, over a finite
interval $(-R, +R)$ with $R \ge a$. Integrating by parts the
derivative term, and subtracting the interchanged expression $n
\leftrightarrow m$ gives
%
%\begin{equation} \label{Ortho}
$$
i(k_n-k_m) \left[E_n E_m\right]_S \,+\, (k_n^2 - k_m^2)N_{n m} = 0
 \eqno{\rm (A.4)}
$$
%\end{equation}
%
with the overlap integral
%
%\begin{equation} \label{Overlap}
$$
N_{n m} = \int_{-R}^{+R}\! dz \,\varepsilon(z)\, E_n(z)\, E_m(z)
 \eqno{\rm (A.5)}
$$
%\end{equation}
%
and a `surface' term, $\left[F\right]_S = F(R) + F(-R)$. Note that
the $R$-dependence of the surface term $\left[E_n E_m\right]_S$
and that of the overlap integral $N_{nm}$ cancel each other
exactly. We choose normalisation of the RS in such a way that this
property is preserved also for $n=m$:
%
%\begin{equation} \label{OrthoNorm}
$$
 N_{ n m}-\frac{\left[E_n E_m\right]_S}{i(k_n + k_m)} = \delta_{n
 m}\,,
 \eqno{\rm (A.6)}
$$
%\end{equation}
so that the RS do not depend on $R$. The normalisation constant is
chosen such that for bound states {(having Re$\,k_n=0$ and
Im$\,k_n>0$)} the standard normalisation $N_{nn}=1$ is restored,
as here the surface term can be made small for large $R$.  A
careful investigation shows that the Wronskian derivative with
respect to $k$ equals $2 k_n$, so that the GF takes the form of
eq.\,(\ref{GFseries}).

\section{Eigenmodes of the perturbed slab in fig.\,1}
The profile of the dielectric constant of the perturbed slab
sketched in the inset to fig.\,1 has the following form
% \begin{equation}
$$
\varepsilon(z)+\Delta\varepsilon(z)=\left\{
 \begin{array}{cr}
1\,, & z<-a\,,\\
\varepsilon_s\,, &-a<z<b\,,\\
\varepsilon_p\,, &b<z<a\,,\\
1\,, & z>a\,.
 \end{array}
 \right.
% \end{equation}
 \eqno{\rm (A.7)}
$$
The eigenmodes are found as solutions of the Maxwell equation
(\ref{ME}) with the BC of outgoing waves and have the following
explicit form
% \begin{equation}
$$
{\cal E}_\nu(z)=\left\{
 \begin{array}{lr}
A e^{-i\varkappa_\nu z}\,, & z<-a\,,\\
B e^{i\sqrt{\varepsilon_s}\varkappa_\nu z}+C e^{-i\sqrt{\varepsilon_s}\varkappa_\nu z}\,, &-a<z<b\,,\\
D e^{i\sqrt{\varepsilon_p}\varkappa_\nu z}+ E e^{-i\sqrt{\varepsilon_p}\varkappa_\nu z}\,, &b<z<a\,,\\
F e^{i\varkappa_\nu z}\,, & z>a\,.
 \end{array}
 \right.
 \eqno{\rm (A.8)}
$$
%\label{EFnu}
% \end{equation}
The six coefficients in eq.\,(A.8)
%\ref{EFnu})
are found (for each $\nu$) using the continuity of the ${\cal
E}_\nu(z)$ and ${\cal E}_\nu'(z)$ on the three boundaries ($z=-a$,
$z=b$, and $z=a$) and the normalisation condition eq.\,(A.6).
%\ref{OrthoNorm}).
The set of six equations following
from the above mentioned BC not only establishes the relations
between the coefficients but also leads to the following secular
equation for the wave numbers $\varkappa_\nu$:
%\begin{equation}
$$
\alpha\beta f(\varkappa_\nu)g(\varkappa_\nu)-1
=\frac{\beta-\alpha}{\alpha\beta-1}\Bigl[\beta
g(\varkappa_\nu)-\alpha f(\varkappa_\nu)\Bigr]\,,
 \eqno{\rm (A.9)}
$$
% \label{EVapp}
%\end{equation}
where
%\begin{equation}
$$
\alpha=\frac{\sqrt{\varepsilon_s}+1}{\sqrt{\varepsilon_s}-1}\,, \
\ \ \beta=\frac{\sqrt{\varepsilon_p}+1}{\sqrt{\varepsilon_p}-1}\,,
 \eqno{\rm (A.10)}
$$
%\end{equation}
 and the functions
$f(k)$ and $g(k)$ are defined as follows:
%\begin{eqnarray}
$$
f(k)=\exp\left[-2i\sqrt{\varepsilon_s} k(a+b)\right]\,,
%\\
 \eqno{\rm (A.11)}
$$
$$
g(k)=\exp\left[-2i\sqrt{\varepsilon_p} k(a-b)\right]\,.
 \eqno{\rm (A.12)}
$$
%\end{eqnarray}

\section{Derivation for the 3D case with spherical symmetry}
We consider a dielectric sphere of radius $a$ with a dielectric
constant of spherical symmetry, $\varepsilon(r)$, embedded into
vacuum. The solutions of Maxwell's equation can be classified into
TE and TM modes, and the angular momentum is a conserved quantity.
The Green's function can be separated into $l$-diagonal
contributions,
%
%\begin{equation}
$$
G(\mathbf{r}, \mathbf{r}'; k) = \frac{1}{r r'} \sum_{l=0}^\infty
G_l(r, r'; k) \frac{2l+1}{4\pi} P_l(\cos\Theta)
 \eqno{\rm (A.13)}
$$
%\end{equation}
%
where $\Theta$ denotes the angle between $\mathbf{r}$ and
$\mathbf{r}'$, and $P_l(x)$ are the Legendre polynomials. For
simplicity, we consider only the TE case {below.}  The GF obeys
the equation
%
%\begin{equation}
$$
\left[\frac{d^2}{dr^2} - \frac{l(l+1)}{r^2}+ k^2\varepsilon(r)
\right]G_{ l}(r, r';k) = \delta(r-r')\,,
 \eqno{\rm (A.14)}
$$
%\end{equation}
%
which differs from the 1D version eq.\,(\ref{GFequ}) only by the
centrifugal term $\propto 1/r^2$. For the homogeneous equation, we
define a solution $f_k(r)$ which is regular at the origin,
%
%\begin{equation}
$$
f_k(r) = r^{l+1} \  \mbox{for} \ r \to 0
 \eqno{\rm (A.15)}
$$
% \label{fk}
%\end{equation}
%
and another one, $g_k(r)$, which has outgoing wave BC:
%
%\begin{equation}
$$
g_k(r) = e^{+ikr} \  \mbox{for} \  r \to \infty \, .
 \eqno{\rm (A.16)}
$$
%\end{equation}
%
As in the 1D case, the GF can be constructed using both solutions,
%
%\begin{equation} \label{GFWron3}
$$
G_{ l}(r,r';k) = \frac{f_k(r_<)\, g_k(r_>)}{W\{f_k,g_k\}} \, .
 \eqno{\rm (A.17)}
$$
%\end{equation}
%
The analytic continuation into the complex $k$ plane has poles
wherever the Wronski determinant has zeroes which defines the wave
numbers of the resonant modes,
%
%\begin{equation}
$$
W\{f_k,g_k\} = 0 \  \mbox{at} \ k = k_n \, .
 \eqno{\rm (A.18)}
$$
%\end{equation}
%
At $k_n$, both solutions $f_k$ and $g_k$ agree up to a factor.
Integrating $G_{ l}(r, r'; k')/(k'-k)$ over a large circle in the
complex $k'$ plane gives zero, but can be decomposed into pole
contributions with the result
%\begin{equation}
$$
G_{ l}(r, r'; k) = \sum_n \frac{f_{k_n}(r)\,
g_{k_n}(r')}{(k-k_n)\, w_n}
 \eqno{\rm (A.19)}
$$
%\end{equation}
%
(note that in 3D, there is no trivial pole at $k = 0$). The
Wronskian derivative, {$w_n=d W/dk|_{k=k_n}$,}
%\revision{X}
%\begin{equation}
%$$
% w_n = \left. \frac{d W\{f_k,g_k\}}{dk} \right|_{k=k_n}  ,
% \eqno{\rm (A.20)}
%$$
%%\end{equation}
%
is calculated integrating Maxwell's equation by parts over the
finite interval $(0, R)$ with $R \ge a$ which gives
%
%\begin{eqnarray} \label{WHatk}
$$
\hspace{-0.5cm} \frac{d W\{f_k,g_k\}}{dk} = f_k(R) \,\frac{ d
g'_k(R)}{dk} - f'_k(R)\, \frac{d g_k(R)}{dk}
$$
$$
\hspace{3cm} + 2 k \int_0^R dr \,\varepsilon(r)\, g_k(r)\, f_k(r)
\, ,
 \eqno{\rm (A.20)}
$$
%\end{eqnarray}
owing to the $k$-independent regular solution eq.\,(A.15)
%\ref{fk})
at $r\to0$. A further simplification is possible due to $R \ge a$
where $\varepsilon(r) \equiv 1$ holds. Here, the outgoing wave
solution depends on $r$ and $k$ only in product form, $g_k(r) =
\tilde{g}(x = kr)$. Then, the first term on the r.h.s. of
eq.\,(A.20)
%\ref{WHatk})
can be rewritten as
%
%\begin{equation}
$$
\left. f_k(R) \left[\tilde{g}'(x) + x \tilde{g}''(x) \right] -
f'_k(R) R \tilde{g}'(x) \right|_{x = kR}\,.
 \eqno{\rm (A.21)}
$$
%\end{equation}
%
At the poles, we have $f_{k_n}(r) = c_n \cdot g_{k_n}(r)$, and can
take $g_n(r) \equiv g_{k_n}(r)$ everywhere,
%
%\begin{eqnarray} \label{Norm3D}
$$
w_n = \left. \tilde{g}(x) \tilde{g}'(x) + x \left[\tilde{g}(x)
\tilde{g}''(x) - (\tilde{g}'(x))^2\right]
\right|_{x = k_n R}
%\nonumber \\
$$
$$
 \hspace{-2.2cm} + 2 k_n \int_0^R dr \,\varepsilon(r) \,g_n^2(r)
 \eqno{\rm (A.22)}
 $$
% \label{wn}
%\end{eqnarray}
in conjunction with
%
%\begin{equation}
$$
G_{ l}(r, r'; k) = \sum_n \frac{g_n(r)\, g_n(r')}{ (k-k_n)\, w_n }
\, .
 \eqno{\rm (A.23)}
$$
% \label{GFapp}
%\end{equation}
%
Equation~(A.22)
%\ref{Norm3D})
can be considered as the proper
normalisation of the resonant modes.  Its l.h.s. is formally
equivalent to their orthogonality (valid for $k_n \neq k_m$),
%\begin{eqnarray}
$$
\hspace{-2.3cm} 0 = \frac{g_n(R) g'_m(R) - g'_n(R)
g_m(R)}{k_m-k_n}
$$
%\nonumber \\
$$
\hspace{1.5cm} + \,(k_n+k_m) \int_0^R dr \,\varepsilon(r)\, g_n(r)
\,g_m(r) \, ,
 \eqno{\rm (A.24)}
$$
%\end{eqnarray}
%
if the limit $k_m \rightarrow k_n$ is performed.


\begin{thebibliography}{0}
\bibitem{Shroedinger26}
  \Name{Schr\"odinger E.}
  \REVIEW{Ann. Physik}{80}{1926}{437}.
\bibitem{Brillouin32}
  \Name{Brillouin L.}
  \REVIEW{J. Phys. Radium}{3}{1932}{373}.
\bibitem{Gamow28}
  \Name{Gamow G.}
  \REVIEW{Z. Phys.}{51}{1928}{204};
  \REVIEW{Z. Phys.}{52}{1929}{510}.
\bibitem{Zeldovich60}
  \Name{Zel'dovich Ya.}
  \REVIEW{Zh. Eksp. Teor. Fiz.}{39}{1960}{776}.
[\REVIEW{Sov. Phys.-JETP}{12}{1961}{542}].
\bibitem{Zeldovich69}
  \Name{Baz' A., Zel'dovich Ya. \and Perelomov A.}
  \Book{Scattering, Reactions and Decay in Nonrelativistic Quantum
Mechanics}
  \Publ{U. S. Department of Commerce, Washington, D. C.}
  \Year{1969}.
\bibitem{Weinstein69}
  \Name{Weinstein L.\,A.}
  \Book{Open Resonators and Open Waveguides}
  \Publ{Golem Press, Boulder, Colorado}
  \Year{1969}.
 \bibitem{More71}
  \Name{More R.\,M.}
  \REVIEW{Phys. Rev. A}{4}{1971}{1782}.
\bibitem{More73}
  \Name{More R.\,M. \and Gerjuoy E.}
  \REVIEW{Phys. Rev. A}{7}{1973}{1288}.
\bibitem{Olshansky79}
  \Name{Olshansky R.}
  \REVIEW{Rev. Mod. Phys.}{51}{1979}{341}.
\bibitem{Tikhodeev02}
%  \Name{Tikhodeev S.\,G. \etal}
  \Name{Tikhodeev S.\,G., Yablonskii A.\,L., Muljarov E.\,A., Gippius N.\,A. \and Ishihara T.}
  \REVIEW{Phys. Rev. B}{66}{2002}{045102}.
\bibitem{Leung94}
%  \Name{Leung P.\,T. \etal}
  \Name{Leung P.\,T., Liu S.\,Y., Tong S.\,S. \and Young K.}
  \REVIEW{Phys. Rev. A}{49}{1994}{3068}.
\bibitem{Leung96}
  \Name{Leung P.\,T. \and Pang K.\,M.}
  \REVIEW{J. Opt. Soc. Am. B}{13}{1996}{805}.
\bibitem{Lai90}
%  \Name{Lai H.\,M. \etal}
  \Name{Lai H.\,M., Leung P.\,T., Young K., Barber P.\,W. \and Hill S.\,C.}
  \REVIEW{Phys. Rev. A}{41}{1990}{5187}.
\bibitem{Dettmann08}
 \Name{Dettmann C.\,P., Morozov G.\,V., Sieber M. \and
H. Waalkens }
%  \Name{Dettmann C.\,P. \etal}
{   \REVIEW{Phys. Rev. A}{80}{2009}{063813}.}
\bibitem{Bang78}
%  \Name{Bang J. \etal}
  \Name{Bang J., Gareev F.\,A., Gizzatkulov M.\,H. \and S.\,A. Goncharov}
  \REVIEW{Nuclear Physics A}{309}{1978}{381}.
\bibitem{Lind93}
  \Name{Lind P.}
  \REVIEW{Phys. Rev. C}{47}{1993}{1903}.
{\bibitem{Aguiar93} \Name{M.\,A.\,M. de Aguiar}
  \REVIEW{Phys. Rev. A}{48}{1993}{2567}.}
\bibitem{Newton60}
  \Name{Newton R.}
  \REVIEW{J. Math. Phys.}{1}{1960}{319}.
{\bibitem{Sukumar90} \Name{C.\,V. Sukumar}
  \REVIEW{Am. J. Phys.}{58}{1990}{561}.}
\bibitem{Bang81}
  \Name{Bang J.  \and Gareev F. A.}
  \REVIEW{Lett. Nuovo Cimento}{32}{1981}{420}.
\bibitem{footnote-QM} A quantum-mechanical analogue of eq.\,(\ref{BWtheory}) can be found in
refs.~\cite{Bang78} and \cite{Lind93}.
\bibitem{Watson86}
  \Name{Watson D.\,K.}
  \REVIEW{Phys. Rev. A}{34}{1986}{1016}.
%\bibitem{note1} Surrounding the inhomogeneity region
%by a sphere is not a limitation of the theory. Generalisation to
%quasi-1D and 2D structures like photonic crystal slabs is
%possible.
\bibitem{note2} This can be shown explicitly expanding the ${\bf E}_n({\bf
r})$ into the complete set of vector spherical
harmonics~\cite{Stratton}.
\bibitem{Stratton}
  \Name{Stratton J.\,A.}
  \Book{Electromagnetic theory}
  \Publ{McGraw-hill Book Company, New York and London}
  \Year{1941}.
\end{thebibliography}
\end{document}